\documentclass[sigconf]{acmart}

\usepackage{float}
\usepackage{xcolor}
\usepackage{multirow}
\usepackage{paralist}
\usepackage{booktabs}
\usepackage{xspace}
\usepackage{subcaption}
\usepackage{balance} 
\definecolor{black}{rgb}{0,0,0}
\definecolor{grey}{rgb}{0.8,0.8,0.8}
\definecolor{red}{rgb}{1,0,0}
\definecolor{green}{rgb}{0,1,0}
\definecolor{darkgreen}{rgb}{0,0.5,0}
\definecolor{darkpurple}{rgb}{0.5,0,0.5}
\definecolor{darkdarkpurple}{rgb}{0.3,0,0.3}
\definecolor{blue}{rgb}{0,0,1}
\definecolor{shadegreen}{rgb}{0.95,1,0.95}
\definecolor{shadeblue}{rgb}{0.95,0.95,1}
\definecolor{shadered}{rgb}{1,0.85,0.85}
\definecolor{shadegrey}{rgb}{0.85,0.85,0.85}
\definecolor{oddRowGrey}{rgb}{0.80,0.80,0.80}
\definecolor{evenRowGrey}{rgb}{0.85,0.85,0.85}

\definecolor{ForestGreen}{rgb}{0.0, 0.66, 0.47}
\definecolor{RubineRed}{rgb}{1.0, 0.0, 0.31}

\newcommand{\green}[1]{{\textcolor{ForestGreen}{{#1}}}}
\newcommand{\blue}[1]{{\textcolor{blue}{{#1}}}}
\newcommand{\red}[1]{{\textcolor{RubineRed}{{#1}}}}

\newtheorem{Problem}{Problem}

\newcommand{\system}{\ensuremath{\mathsf{Machop}}\xspace}
\newcommand{\ditto}{\textsf{Ditto}\xspace}
\newcommand{\longf}{\textsf{Longformer}\xspace}
\newcommand{\dm}{\textsf{Deep Matcher}\xspace}

\newcommand{\attr}{\mathsf{attr}}
\newcommand{\val}{\mathsf{val}}

\usepackage[shortlabels]{enumitem}
\setlist[itemize]{leftmargin=*}
\setlist[enumerate]{leftmargin=*}

\usepackage[linesnumbered,ruled,vlined]{algorithm2e}

\usepackage{dsfont}

\usepackage{multirow}
\usepackage[font=small,skip=3pt]{caption}

\AtBeginDocument{%
  \providecommand\BibTeX{{%
    \normalfont B\kern-0.5em{\scshape i\kern-0.25em b}\kern-0.8em\TeX}}}

\copyrightyear{2022}
\acmYear{2022}
\setcopyright{acmcopyright}\acmConference[aiDM'22 ]{Exploiting Artificial Intelligence Techniques for Data}{June 17, 2022}{Philadelphia, PA, USA}
\acmBooktitle{Exploiting Artificial Intelligence Techniques for Data (aiDM'22 ), June 17, 2022, Philadelphia, PA, USA}
\acmPrice{15.00}
\acmDOI{10.1145/3533702.3534910}
\acmISBN{978-1-4503-9377-5/22/06}

\begin{document}

\title{\system: an End-to-End Generalized Entity Matching Framework}

\author{Jin Wang}
\email{jin@megagon.ai}
\affiliation{%
  \institution{Megagon Labs}
  \country{United States}
}

\author{Yuliang Li}
\email{yuliang@megagon.ai}
\affiliation{%
  \institution{Megagon Labs}
  \country{United States}
}

\author{Wataru Hirota}
\email{wataru@megagon.ai}
\affiliation{%
  \institution{Megagon Labs}
  \country{United States}
}

\author{Eser Kandogan}
\email{eser@megagon.ai}
\affiliation{%
  \institution{Megagon Labs}
  \country{United States}
}
\renewcommand{\shortauthors}{Wang et al.}

\begin{abstract}
Real-world applications frequently seek to solve a general form of the
Entity Matching (EM) problem to find associated entities.
Such scenarios include matching jobs to candidates in job targeting,
matching students with courses in online education, matching products with user reviews on e-commercial websites, and beyond.
These tasks impose new requirements such as matching data entries with diverse formats or having a flexible and semantics-rich matching definition, which are beyond the current EM task formulation or approaches.

In this paper, we introduce the problem of Generalized Entity Matching (GEM) that satisfies these practical requirements and presents an end-to-end pipeline \system as the solution.
\system allows end users to define new matching tasks from scratch and apply them to new domains in a step-by-step manner. 
\system casts the GEM problem as sequence pair classification so as to utilize the language understanding capability of Transformers-based language models (LMs) such as BERT.
Moreover, it features a novel external knowledge injection approach with structure-aware pooling methods that allow domain experts to guide the LM to focus on the key matching information thus further contributing to the overall performance.
Our experiments and case studies on real-world datasets from a popular recruiting platform show a significant 17.1\% gain in $F_1$ score against state-of-the-art methods along with meaningful matching results that are human understandable.

\end{abstract}

\maketitle

\section{Introduction} \label{sec:intro}

Entity Matching (EM), also known as entity resolution, record linkage, reference reconciliation, and duplicate detection, refers to the problem of identifying pairs of  data entries representing the same real-world entity~\cite{fellegi69,DBLP:journals/tkde/Christen12,DBLP:books/daglib/0029346,DBLP:journals/pvldb/KondaDCDABLPZNP16,li2021deep}.
As one of the fundamental problems in data integration, EM has a wide range of applications including data cleaning, knowledge base construction, clustering, and search.
More recently, deep learning techniques~\cite{DBLP:conf/edbt/Thirumuruganathan20,DBLP:conf/edbt/BrunnerS20,ditto2021}, 
especially pre-trained language models (LMs)~\cite{DBLP:conf/naacl/DevlinCLT19,DBLP:journals/corr/abs-1907-11692,DBLP:journals/corr/abs-1910-01108,DBLP:conf/iclr/LanCGGSS20},
have achieved promising results on EM tasks.

In real-world applications, many practical scenarios could potentially benefit from a generalized formulation of EM problem, 
where entities may be in a variety of formats and the association between them can go beyond equality-based matching. 
Examples include matching jobs to candidates in job targeting, matching students with courses in online education, matching products with user reviews on e-commercial websites etc.

\smallskip
\noindent
\textbf{Example and Challenges. } 
Next, we will use the application of job targeting to illustrate the challenges.
There are two important tasks in job targeting, namely job-job matching and job-resume matching.
While there are many previous studies on job targeting mainly leveraging contextual~\cite{DBLP:conf/kdd/LiAHS16,DBLP:conf/ijcai/ShenZZXMX18,DBLP:conf/kdd/YanLSZZ019} and historical information~\cite{DBLP:conf/sigir/QinZXZJCX18,DBLP:conf/cikm/JiangYWXL20,DBLP:conf/cikm/LeHSZ0019,DBLP:conf/cikm/Bian0ZZHSZW20}, 
in this paper we focus on formulating the task as entity matching
on the textual content of the entities. 
% \yuliang{modified a bit. Shall we also say that ``Although extra features such as
% search logs or application history are not considered, they can be captured by our framework by an extension.''}
% the textual features themselves without considering additional features.

\begin{figure}[!ht]
    \centering
    \includegraphics[width=0.48\textwidth]{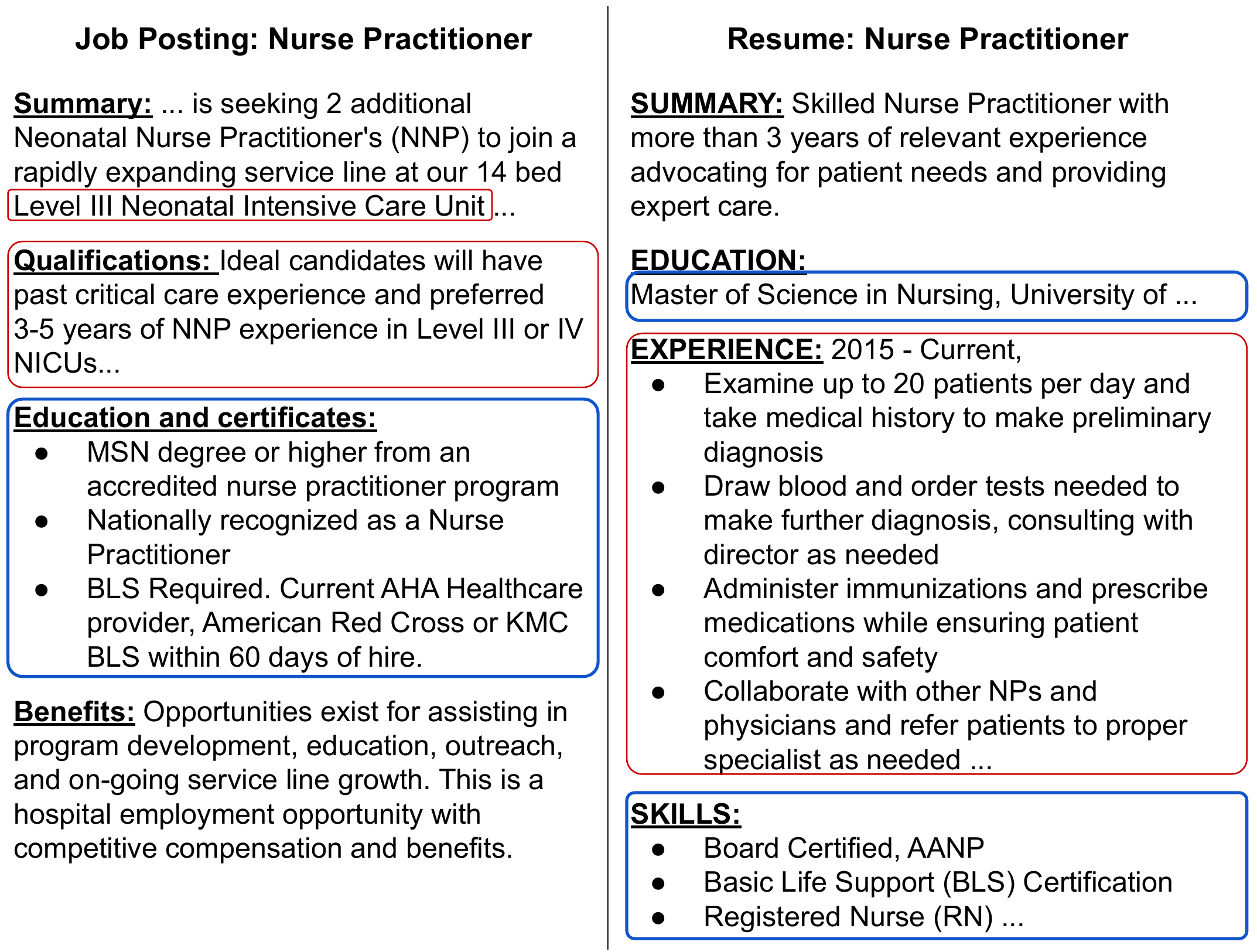}
    \caption{An instance of generalized EM in job-resume matching.
    While the resume matches with the job posting on the required skills and education, the job requires specific work experiences in NICUs. As such the resume should be disqualified. 
    The example illustrates the need for both improved language understanding and structural domain knowledge to make the right match. }
    \label{fig:example}
\end{figure}

Figure~\ref{fig:example} shows real, anonymized data from job postings and resumes 
from the job targeting platform of 
Indeed.com.
The classic EM problem definition faces with two major challenges in supporting such applications:

%Figure~\ref{fig:example} illustrates an instance of the job-resume matching task. 
%Both entries (anonymized) are from real datasets of job postings and resumes of Company X's job targeting application~\footnote{The company name in the rest of the paper is replaced with ``Company X'' due to the double-blind policy.}. 
%The goal is to identify and match qualified candidates with job positions. 

First, the data entries from different data collections are in \emph{heterogeneous data formats}, while solutions to the classic EM problem typically assume the two collections to be structured and have the same or aligned schema.
However, this assumption is hard to be satisfied in real applications.
In this example, after pre-processing from raw data (PDF or HTML), the resume and job entries are in either unstructured text or semi-structured format that contains nested attributes or lists. 
As such, it is unlikely for them to have perfectly aligned attributes to apply classic EM solutions.

Second, classic EM solutions primarily focus on matching data entries that refer to a real-world entity, while the definition of matching might be more general in real applications.
In the example of job-resume matching shown in Figure~\ref{fig:example}, the entities to match are a resume(candidate) and a job posting(job); and the definition of matching is whether a candidate's resume is qualified for a jobs. 
Capturing such relations is challenging even to the most advanced EM or NLP techniques. 
As shown in this example, although the two data entries share the same title of ``nurse practitioner'' and have high textual similarity across the content, they should not be matched because the job posting explicitly requires qualification with a working experience in a specific type of intensive care unit.

To successfully handle this task, advanced language understanding is required to match each qualification item on the job posting with an item of education, skill, or experience in the resume.
Besides, an \emph{improved understanding of the entry structure} is also needed to guide the underlying model to identify the important sections that need to be matched against each other.
In this case, the ``qualification'' of the job posting and the ``experience'' of the resume are important and need to be matched (circled in \red{red}). 
Similarly, ``education and certificates''  needs to match with the ``education'' and ``skills'' sections of the candidate (circled in \blue{blue}). 
Other contents such as the ``benefits'' and ``summary'' are deemed less important in matching qualified candidates to jobs.

\begin{figure*}[!t]
    \centering
    \includegraphics[width=0.9\textwidth]{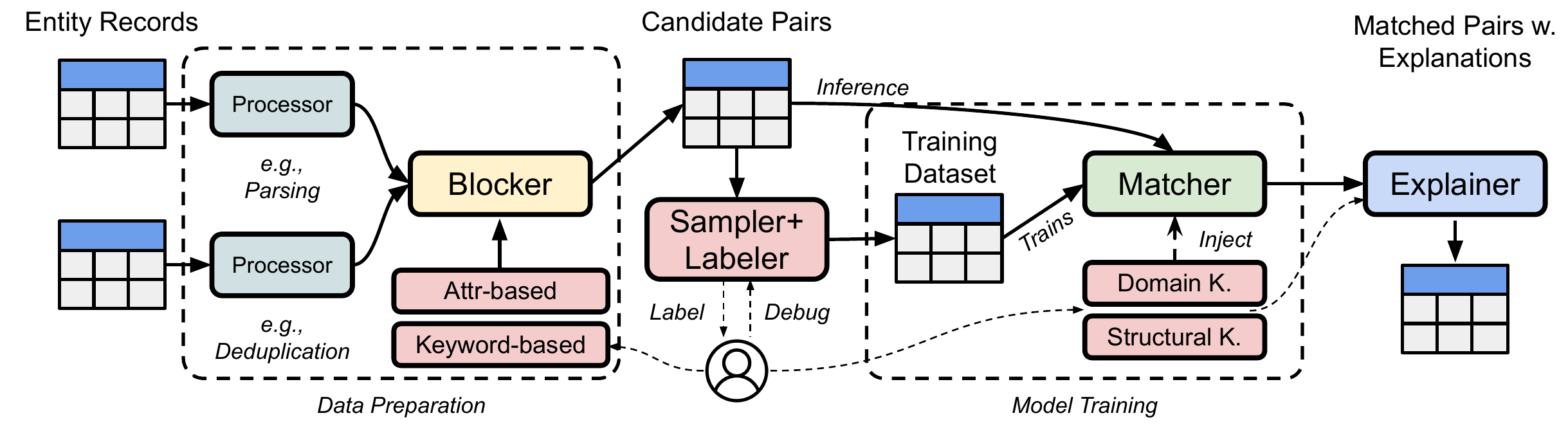}
    \caption{The \system pipeline}
    \label{fig:gem}\vspace*{-1mm}
\end{figure*}

% \smallskip
\noindent\textbf{The \system pipeline. }
In this paper, we formulate the problem of \emph{Generalized Entity Matching} (GEM) to accommodate a wider range of matching applications.
The new problem formulation explicitly allows matching to happen between data entries with diverse data formats (structured, semi-structured, and unstructured)
and with a customizable matching definition. % general binary association relation. 
To this end, we propose \system, an end-to-end pipeline as the solution (Figure \ref{fig:gem}).
The \system pipeline enables developers to build generalized matching solutions from scratch in a step-by-step manner. 
Starting from raw input data collections, \system applies (customizable) pre-processing and blocking heuristics to create candidate pairs to be matched. 
Based on that, it allows end users to semi-automatically create and debug the labels over candidate pairs.
In the matching step, \system provides two effective model architectures based on fine-tuning pre-trained LMs, e.g. BERT~\cite{DBLP:conf/naacl/DevlinCLT19}, to predict the matched entities.
As such, with its broader utility, \system has good potential to be applied in a wider range of real applications.

In summary, this paper makes the following contributions:
\begin{compactitem}
\item We introduce the Generalized Entity Matching (GEM) problem by allowing (i) two entities can have different data formats; and (ii) the definition of matching can be more general and customized based on different real applications.
\item We present \system, an end-to-end pipeline as a practical solution to the GEM problem.
With its customizable modules for creating high-quality training/test datasets in a semi-automatic manner, it has broad utility to become a key tool for end users.
We demonstrate the viability of \system by deploying it on two important tasks from job targeting: job-job and job-resume matching.
\item We propose novel model architectures based on pre-trained LMs to predict the matching results precisely. 
The new architectures feature a novel \emph{structure-aware pooling layer} enabling the models to capture the rich structural knowledge within the entities.
It further allows domain experts to transform long text fields into (semi-)structured attribute-value pairs via domain knowledge injection to improve the matching quality.
\item We evaluate \system on real-world job targeting datasets.
Experimental results show that the proposed matching models in \system outperform state-of-the-art EM methods by over 17\% $F_1$ score.
Our case study also demonstrates its capability to perform matching that requires nontrivial domain and language understanding ability while being able to generate human-readable explanations to each matching decision.
\end{compactitem}

The rest of the paper is organized as follows. 
Section~\ref{sec:overview} overviews the \system pipeline. 
Section~\ref{sec:model} presents the model architectures and 
the domain knowledge injection optimization.
Section~\ref{sec:exp} reports the experiment results and Section~\ref{sec:casestudy} presents case studies of applying \system to real applications. 
Section~\ref{sec:related} surveys the related work and Section~\ref{sec:conclude} concludes the paper.

\section{The \system matching pipeline} \label{sec:overview}

Next, we present the problem definition and \system's architecture.

\subsection{The GEM problem formulation}\label{subsec:problem}

Given two collections of entities $E_A$ and $E_B$, the \emph{Entity Matching} (EM) problem aims at identifying all pairs of 
entity entries $\langle e_a, e_b \rangle$ for $e_a \in E_A$, $e_b \in E_B$ that refer to the same real-world entity.
Classic settings of EM typically assume that they are two relational tables having the same schema. 
Namely, there exists a relation $R(\attr_1, \dots, \attr_n)$ with $n$ attributes such that all entries in 
$E_A$ and $E_B$ are elements of $R$.

To formulate the GEM problem, we extend the definition as follows. 
We assume two base data types of \emph{number} and \emph{string}~\footnote{We also include the list/array type but we omit it for sake of space.}. 
An entity entry $e$ is a named tuple (i.e., a key-value collection) with a set of attributes  $\{\attr_1, \dots, \attr_n\}$ where $\attr_i$ is the attribute name and $e.\attr_i$ denotes the attribute value.
Each value $e.\attr_i$ can be of either the base type or an entity entry itself. 
In other words, it allows $e$ to be semi-structured data with nested attributes such as JSON. 
We define an entity collection $E$ (or collection for short) to be a set of entity entries.
When $n > 1$ or $E$ contains a entry with nested attributes, we call it a semi-structured collection; $E$ is structured (or relational) otherwise.
When $n = 1$ and the only attribute $e.\attr_1$ is of the string type, we call the entry $e$ ($E$) an unstructured entry (collection).
% \eser{JSON also has an array value. Do you want to explicitly call that we don't handle array value, or ?}\yuliang{we mentioned that in a footnote that we also support lists}

We also relax the definition of matching which is ``two entities refer to the same real-world entity'' in classic EM as follows.
Given two entities $e_a$ and $e_b$, we say that they match with each other in GEM when a predefined condition $R_{\mathsf{match}}(e_a, e_b)$ is satisfied.
For example, in an e-commercial website, if $E_A$ is a structured table with entities of product information and $E_B$ is a text corpus of product review, $R_{\mathsf{match}}(e_a, e_b)$ can be defined as $e_b$ being a valid review of product $e_a$. 

Following the above discussion, we can then formally define the GEM problem:

\begin{Problem}[GEM]
	Given two entity collections $E_A$ and $E_B$ where each of them can be structured, semi-structured, or unstructured, a binary matching relation 
	$R_{\mathsf{match}}$, it aims at finding all entry pairs $\langle e_a, e_b \rangle$ for $e_a \in E_A$, $e_b \in E_B$ where
	$e_a$ and $e_b$ satisfy $R_{\mathsf{match}}(e_a, e_b)$.
\end{Problem}

\subsection{Example application scenarios of GEM}\label{subsec-job}

In this paper, we demonstrate the versatility of GEM 
using the job targeting application as an example.
Specifically, we formulate two important tasks in job targeting, job-job matching and job-resume matching, as tasks of GEM.
Both tasks have real product impact in popular job targeting platforms
such as Indeed.com. 
For job-job matching, the goal is to find pairs of job postings similar to each other
from a collection of jobs. Such pairs can later be used in downstream tasks
such as job recommendation or de-duplication of job search results.
The goal of job-resume matching is to identify resumes that are qualified for certain jobs. 
The matching results can then be used to invite qualified candidates to apply for the posted jobs.

\begin{figure}[h!t]
	\centering
	\includegraphics[width=0.48\textwidth]{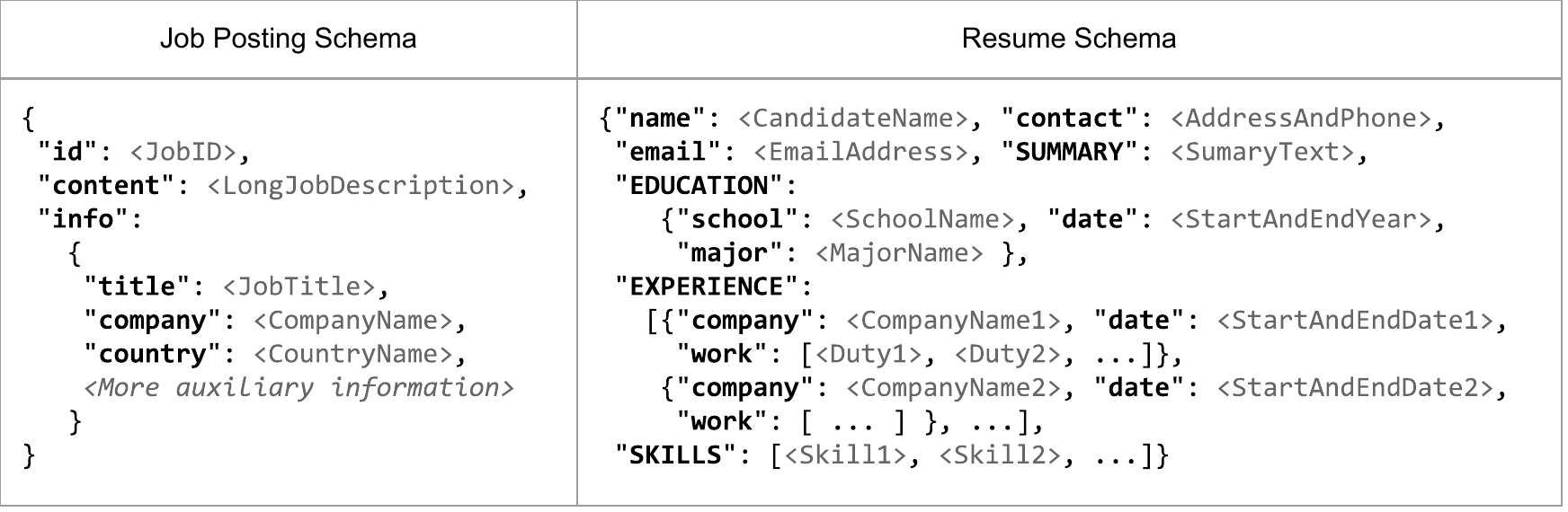}
	\caption{\small The job posting and resume schema. The resume data contains more structured data while
		most useful information of the job posting is a long textual document stored in the ``content'' attribute.}
	\label{fig:schema}
\end{figure}

Figure~\ref{fig:schema} shows the schema of the job and resume entries after they are pre-processed from the original PDF or HTML files~\footnote{Due to privacy concern, the schema here is not exactly the same with that in the real product pipeline}. 
The details are as following:

A job posting record consists of the following fields: \emph{id} is the internal id of the record; \emph{content} is a long document that includes the details of the job description; \emph{info} contains 4 attributes: the \emph{title} of a job posting, the name of \emph{company}, its \emph{location} and the \emph{category} in the format of internal ids. 
As the useful information for matching is just the title and content, we regard the job posting record as an unstructured document which is the concatenation of \emph{title} and \emph{content}.

A resume record consists of three categories of information: the personal information, the \emph{education}, the \emph{experience} and the \emph{skills} of a candidate. 
Each category is further split into finer-grained nested attributes. 
For example, the education field has the attributes of school name, duration, major, and degree etc.
All attributes except the personal information are used for matching.
Thus we regard a resume as a semi-structured record with nested attributes.

Therefore, we formulate above two tasks as the GEM problem: \emph{job-job matching is to match two entries in unstructured text; while job-resume matching is to match a semi-structured entry and an unstructured entry without a clearly aligned schema.}
Moreover, we also need GEM's customizable notions of matching
as both matching conditions ``two jobs being related the each other'' 
and ``the resume is qualified for the job'' are beyond the simple
equality relation.

Furthermore, by formulating the two tasks as GEM,
we enable a data science workflow in a style similar to that of classic EM, e.g.
Magellan~\cite{DBLP:journals/pvldb/KondaDCDABLPZNP16}, 
for building matching solutions from scratch in a step-by-step manner.
Indeed, as we describe next,
our general-purpose \system framework allows end users to build GEM pipelines starting from raw unlabeled entity entries, making it easy to be deployed to a wide range of real applications.

\subsection{Overall architecture}\label{subsec:overall}

Figure~\ref{fig:gem} shows the overall architecture of the \system pipeline.
The details of key steps in it are described as following.

\subsubsection{Data Preparation}

The goal of data preparation is to pre-process raw entity entries and apply blocking to generate candidate entry pairs for model training and matching.
It consists of two components: Processor and Blocker.

\smallskip
\noindent\textbf{Processor}\hspace{.5em} 
Given input sets of raw entries, \system first needs to parse and clean the datasets. 
To this end, it provides a set of off-the-shelf 
data processors for this purpose (can be user-defined as well).
For example, for resume entries, we need to convert the resumes to JSON format using a PDF/HTML parser.
The job posting dataset contains a significant number of duplicate entries  (e.g., the same job posted in multiple locations) and spam information (e.g. with nothing but a URL of the employee's website). 
We apply some rule-based processors, e.g. using n-gram, to keep only 1 entry for each group of duplicates.

\smallskip
\noindent\textbf{Blocker}\hspace{.5em} 
The Blocker performs the blocking and filtering mechanisms to avoid the quadratic  number of pairwise comparisons by only keeping candidate pairs that are likely to match. 
For this purpose, the end user needs to specify a \emph{blocking function}, which is typically a fast and rule-based heuristics that keeps most of the matching entries (e.g., high recall is desirable) while resulting in a manageable number of candidate pairs to be matched. 
\system supports a set of widely used blocking functions.
For example, in job-job matching, we apply the exact match rule to the title field
and Q-gram blocking~\cite{DBLP:journals/csur/PapadakisSTP20} to refine the 
resulting pairs.
For job-resume matching, we follow a similar pattern to apply keyword matching rules between the job and position titles attributes.
Note that the end-users are able to inspect the quality of 
the candidate pairs and adjust the blocking strategy to 
ensure that the recall is high enough.
\system supports a standard set of existing blocking strategies 
and how to design more effective ones is beyond the scope of this paper.
\subsubsection{Dataset Creation} 

It requires a high-quality dataset with labeled candidate pairs to train a high-quality pairwise matching model.
After acquiring candidate pairs from the Data Preparation step, the domain expert is then responsible for creating the labels.
In \system, it is realized by the \emph{Dataset Creation} step.
The labeler component in this module provides a user-friendly interface for domain experts to manually label the candidate pairs.
While labeling is a time-consuming process that cannot be skipped, \system provides an optional sampler component for creating negative instances in addition to the manually obtained ones among which a large portion is positive.
For example, in job-job matching, we can sample pairs that have no keyword overlap in their title attributes which implies that they might not probably be matched.
The domain expert can also inspect the quality of labels and debug the blocking and sampling modules as previous EM pipelines did~\cite{DBLP:journals/pvldb/KondaDCDABLPZNP16}.

\subsubsection{Model Training} 

The labeled pairs obtained above are then regarded as training instances in the \emph{Model Training} step.
We propose the matcher based on pre-trained language models (LMs) such as BERT~\cite{DBLP:conf/naacl/DevlinCLT19}, which have shown great effectiveness for EM applications~\cite{DBLP:conf/edbt/BrunnerS20,ditto2021}.
Meanwhile, in GEM applications such as job-resume matching, 
there exists rich structural information within the entity entries
that is useful for matching but not fully utilized by the LMs.
\system's matchers capture such knowledge by
performing an attribute-wise alignment and pooling operator.
For unstructured texts where the structural knowledge exists in hidden sub-structure of the text, \system allows domain experts to \emph{inject domain knowledge} to
long text fields and turn them into meaningful attribute-value pairs.
We provide details of the matcher next in Section~\ref{sec:model}.

\section{The \system matcher} \label{sec:model}

% Next, we introduce the model architectures of \system's matchers and how \system trains the models.

\subsection{Serialization}\label{subsec:seq}

To leverage pre-trained language models (LMs) for matching, \system first converts the
structured or semi-structured input data into sequences of tokens. 
However, we would like the serialization process to 
retain as much structural information as possible to be captured by the LMs.
To achieve this goal, we extend the serialization method proposed in \ditto~\cite{ditto2021} to both structured and semi-structured entries. 

For structured tables, a data entry with $n$ attributes can be denoted as $e = \{ {\attr}_i, {\val}_i\}_{i \in [1,n]}$, where $\attr_i$ is the attribute name and $\val_i$ is the attribute value of the $i$-th attribute, respectively.
Then the serialization is denoted as 
\begin{center}
\small	{\em serialize(e)}: {\tt [COL]}\ $\attr_1$\ {\tt [VAL]}\ $\val_1$ $\ldots$ {\tt [COL]}\ $\attr_n$\ {\tt [VAL]}\ $\val_n$, 
\end{center}
where \textsf{[COL]} and \textsf{[VAL]} are two special tags indicating the start position of attribute names and values respectively.

The semi-structured collection can be serialized in a similar way, where (i) for nested attributes, we recursively add the \textsf{[COL]} and \textsf{[VAL]}  tags along with attribute names and values at each level, and (ii) for list-valued attributes, we concatenate the elements in the list into a string separated by space.
Given the example of resume in Figure~\ref{fig:schema}, we serialize it as:
\begin{center}
	\small 
	[COL] Name [VAL] <candidate name> [COL] Contact [VAL] <candidate contact> ... [COL] Education [VAL] [COL] School [VAL] <school name> [COL] Date [VAL] <StartAndEndYear> ... [COL] Experience [VAL] [COL] Company [VAL] <companyName1> [COL] Date [VAL] <StartAndEndDate1> [COL] Work [VAL] <concatenate work duties> ... [COL] Skills [VAL] <concatenate skills>
\end{center}

For LMs such as BERT that require serialized sequence pairs as input, we concatenate a pair of entities $\langle e, e'\rangle$ 
with a separator token \textsf{[SEP]}. 
We also insert the special token \textsf{[CLS]} necessary for the LM to encode
the entity pair into a feature vector which will be fed into fully connected layers for classification:
\begin{center}
\small
	{\tt [CLS]} {\em serialize(e)} {\tt [SEP]} {\em serialize(e')} 
\end{center}  

Note that the proposed method provides a general serialization scheme for  any nested data such as structured or semi-structured entries.
Besides, the \system framework also allows customized serialization methods for more
complex structures such as graphs~\cite{DBLP:journals/corr/abs-2105-02605}.

\subsection{External knowledge injection}\label{subsec:classifier}

A major challenge in devising effective matchers for the GEM problem is to identify matching sub-structures for heterogeneous data formats.
For attributes with long text such as the ``content'' field in job postings, it is rather difficult to obtain signals for matching even with the help of pre-trained LM due to the limitation of the maximum sequence length.
For example, BERT allows at most 512 sub-word tokens in its input sequence~\cite{DBLP:conf/naacl/DevlinCLT19}.
while real-world entity entries such as job postings or resumes can easily surpass the limit. 
As shown in Table \ref{tbl:datastat},  the job-resume pairs have 727.5 tokens in average which will exceed the 512 limit after being converted into sub-word tokens.
The default option for dealing with long sequences is to truncate the input entries. However, unlike text data,
the key matching information for entity entries might not locate at the beginning thus
a simple truncation can drop important pieces and lead to performance degradation.

To address this issue, \system allows domain experts to inject external knowledge that guides the matcher to select the most informative parts from the entity entries
to keep and pay attention to. 
This is realized by the external knowledge provided by domain experts in the form of a \emph{sentence-topic classifier}.
For job postings, the classifier assigns to each sentence a label from the set
$$\mathsf{\{Qualification, Benefit, Duty, Time, Location, Company\}}$$
which represent the 6 most frequently covered topics in job posting content identified by domain experts. 
There is an additional class ``\textsf{None}'' which represents irrelevant text (e.g., how to apply, email).

Given the classifier, \system enriches the structure of the input by adding a new string-type attribute for each topic.
The value of the attribute consists of the concatenation of all text classified under the topic.
The text classified as ``\textsf{None}'' is removed from the model input to save space.
By doing so, \system turns long text fields such as ``content'' in job postings into structured
attribute-value pairs to retain information of each topic without dropping it entirely.

Note that this technique is general and can be easily extended to new domains.
The domain experts simply need to define the topic classes and provide a few training examples that are easy to acquire 
to train such classifiers (the accuracy does not need to be high).
In this work, we train the classifier by fine-tuning the DistilBERT model~\cite{DBLP:journals/corr/abs-1910-01108} 
using 3k labeled sentences.

More details for the domain knowledge injection module introduced above is as following: 
Given an unstructured document, i.e. the concatenation title and content attributes from a job posting, recall that our goal is to convert the unstructured text into a structured object which provides richer matching information for the underlying model.
To achieve this goal, we formulate this task as sentence-level topic classification that assigns sentences of the documents
into topics such as ``benefit'', ``company'', ``duty'', etc. for job descriptions.

In our implementation, we use the spaCy library to split 
each document into a set of sentences. We provide a rule-based
and a learning-based method to obtain the sentence classifier. 
The rule-based method associates each topic with a set of keywords,
e.g., $\{$``insurance'', ``salary'', ``wage'', $\dots \}$ for the \textsf{benefit}
class. We assign a sentence with the class if it contains any of the keywords.
As shown in our ablation analysis (Section \ref{subsec:ablation}),
this rule-based method leads to a 2\% to 14\% F1 score 
improvement to the sequenced models.

For the learning-based method,
we obtain the sentence classifier by fine-tuning
a DistilBert model on a corpus with 3550 sentences
using the Transformer~\cite{DBLP:conf/emnlp/WolfDSCDMCRLFDS20} library.
These sentences come from the same corpus of job postings and are manually labeled by domain experts from Indeed.com.
The labels include the six categories mentioned in Section~\ref{subsec:classifier} as well as ``None''. 
We randomly split the corpus into the training, validation and test sets 
according the 3:1:1 ratio. 
The classifier is trained on the default set of hyper-parameters: 
the learning rate is set to $3 \times 10^{-5}$, the maximum sentence length is 256 and the batch size is 64.
Finally, the model we used as the classifier achieves
an accuracy of 82.1\% and an $F_1$ score of 74.8\%, which are sufficient in categorizing all sentences 
from the job posting dataset. 

Although labeling effort is required, the learning-based approach achieved better results compared to the rule-based approach.
According to Section \ref{subsec:ablation}, the matching model achieves 1.4\% to 6\% improvement by switching from the rule-based classifier to the learning-based classifier.

\begin{figure}[ht]
\vspace{-0.5em}
	\centering
	\begin{center}
        \begin{subfigure}[b]{0.23\textwidth}
            \centering
            \includegraphics[width=\textwidth]{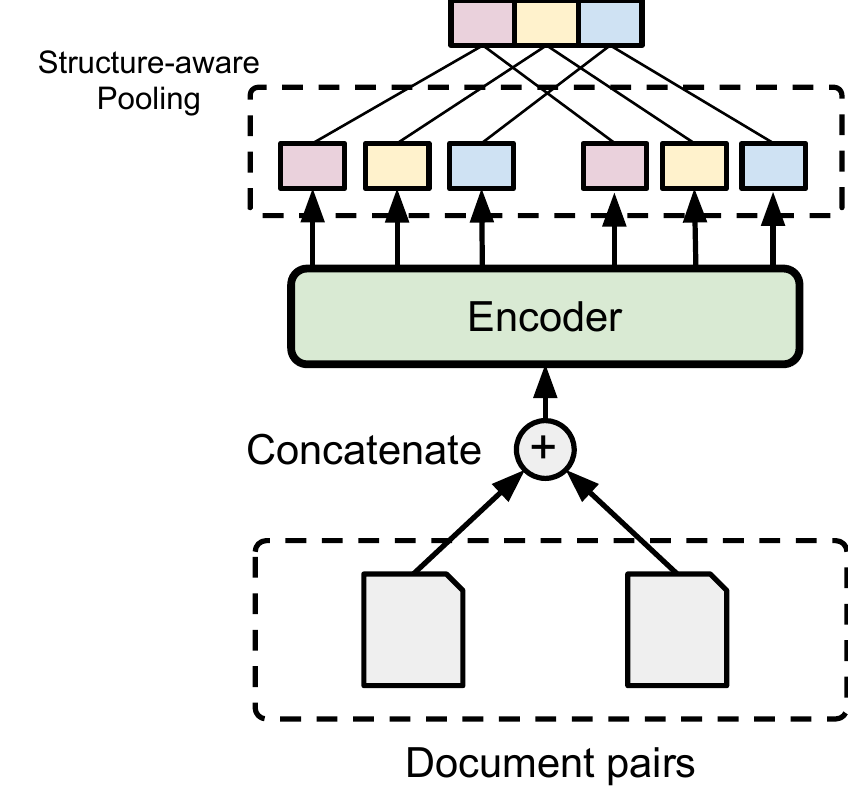}
            \caption{\small{Architecture I: Sequenced}}
            \label{fig:y equals x}
        \end{subfigure}
        \begin{subfigure}[b]{0.23\textwidth}
            \centering
            \includegraphics[width=\textwidth]{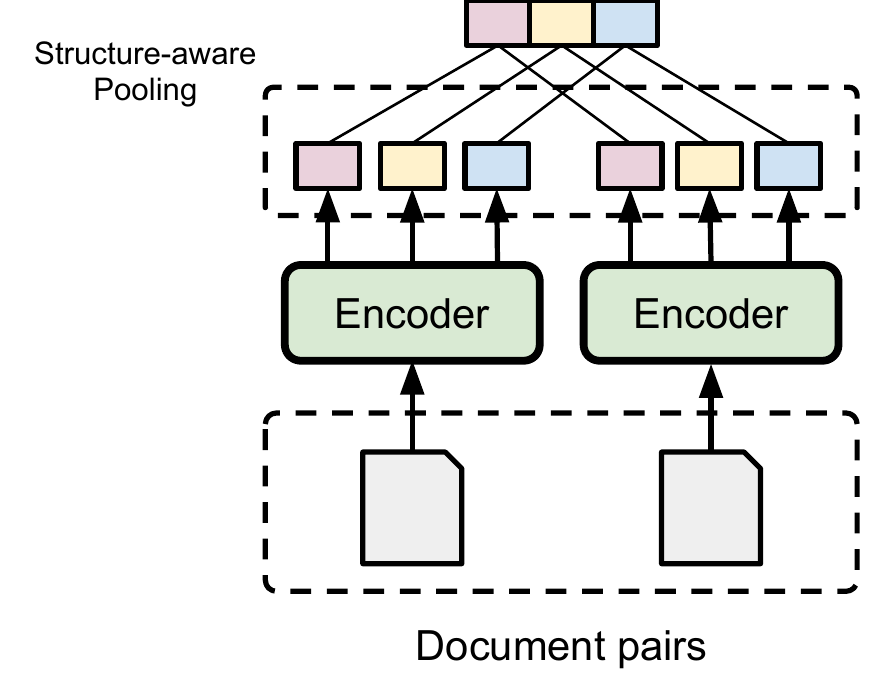}
            \caption{\small{Architecture II: Siamese}}
            \label{fig:three sin x}
        \end{subfigure}	
	\end{center}
	\caption{\small Architectures of \system's matcher. The structure-aware pooling layers aggregate contextualized attribute representations 
	(colored boxes) to leverage the structural or domain knowledge.}\label{fig:arc}
\vspace{-1.5em}
\end{figure}

\subsection{Structure-aware pooling layer}\label{subsec:tspl}

To better utilize the existing structure of the input sequences and make attribute-wise alignment, 
we propose a novel \emph{structure-aware pooling layer} on top of the encoder.
Instead of using the output from the \texttt{[CLS]} token for classification,
the pooling layer leverages the \emph{contextualized attribute representations} of the encoded entity.
We obtain these representations by taking the output of the first token of each encoded attribute~\footnote{This can be realized by replacing the special tokens [COL] with ones that can help distinguish different attributes in serialization, e.g. [DUTY].}.
These output vectors likely provide contextualized attribute-wise representations
since each of them is computed based on the attribute itself and the entire sequence as context 
via the self-attention mechanism of the LM~\cite{DBLP:conf/nips/VaswaniSPUJGKP17}.

The structure-aware pooling layer further leverages alignment information from the input data whenever it is available.
Based on whether the two entries share the same schema, we devise different matching mechanisms in the structure-aware pooling layer.
When two entities have a homogeneous schema, e.g. in the job-job matching task, we leverage the \emph{pairwise similarity} of the aligned attribute pairs as features for matching.
Specifically, for a given set $\{\attr_1, \dots, \attr_n\}$ of aligned attributes (specified by domain experts)
in an entity pair $(e_a, e_b)$,
let $v_i^a$ and $v_i^b$ be the representation of $\attr_i$ in $e_a$ and $e_b$ respectively.
Let $\odot$ be the element-wise product between two vectors and $ \oplus$ be the vector concatenation operation.
The output of the pooling operation is
\begin{equation}\label{equ-poolout1}
	{\mathsf{Pooling\_homo}}(e_a, e_b) = \oplus_{i \in [1,n]} v_i^a \odot v_i^b .
\end{equation}

Meanwhile, when two entities have a heterogeneous schema, e.g. in the job-resume matching task, we let the model \emph{find the most relevant} attributes / features
by applying an attribute-wise max-pooling operator in the structure-aware pooling layer. 
Specifically, let $v_i^a$ be the representation of $\attr_i$ of entity $e_a$. We compute the maximum
matching score for each dimension of $v_i^a$ across all attribute representations $v_j^b$ in entity $e_b$. 
We then concatenate all max matching scores to obtain the features for classification. Formally,
given two entities represented by vectors $\{v_1^a, \dots, v_n^a\}$ and $\{v_1^b, \dots, v_m^b\}$ respectively,
the pooling operator computes
\begin{equation}\label{equ-poolout2}
	{\mathsf{Pooling\_heter}}(e_a, e_b) = \oplus_{i \in [1,n]} \max_{j \in [1,m]} (v_i^a \odot v_j^b) 
\end{equation}
where $\max$ takes the element-wise maximum of the input vectors.
The output of $\mathsf{Pooling\_homo}$ and $\mathsf{Pooling\_heter}$ are then fed into the output layer to compute
the predicted labels and back-propagates to update the model's trainable parameters.

\subsection{Two Model Architectures}\label{subsec:arcs}

To employ pre-trained LMs as encoders to obtain high-quality representations, we proposed two model architectures: \emph{Sequenced} and \emph{Siamese} (Figure~\ref{fig:arc}). 
The Sequenced model has achieved promising results in sequence pair classification 
tasks in many previous studies~\cite{DBLP:conf/naacl/DevlinCLT19,DBLP:journals/corr/abs-1907-11692}.
Although previous work~\cite{DBLP:conf/emnlp/ReimersG19} shows that the Siamese architecture might not perform as well as the Sequenced architecture with pre-trained LM as encoder, it has the potential benefit of learning entity representations that can be pre-computed and used in downstream tasks with efficiency constraints such as in entity similarity search.
Both models consist of four layers: the input layer, encoding layer, structure-aware pooling layer, and output layer.
% \eser{Give some intuition of why there are two models, what do you expect each model to do, if there is a specific case that one performs better than other. What is the intuition? E.g. we expect sequence model to perform better when matching entities of the same type or whatever it is} \yuliang{TODO move the discussion up here.}

In the sequenced architecture the input layer consists of the serialization and domain knowledge injection modules, 
which turn input entity entries into a serialized sequence of token IDs (Figure~\ref{fig:y equals x}). 
Same as the sequence pair classification tasks with pre-trained LMs,
the sequenced architecture concatenates the two serialized entity entries (e.g., a job posting and a resume)
and then feed them into the encoder.
The structure-aware pooling layer is designed as above for different tasks.
The output layer is a linear layer that will be fine-tuned together with the encoding layer
during the training process with a softmax function to perform the binary matching prediction.

In the Siamese Network~\cite{DBLP:conf/eccv/BertinettoVHVT16}, instead of encoding the pairs of entries with a single LM, 
the encoding layer encodes the two entity entries separately using the same encoder (Figure~\ref{fig:three sin x}).
The output of the encoders is then sent to the structure-aware pooling layer for further matching and aggregation.

Regarding the structure-aware pooling layer, we employ a similar idea with the sequenced architecture introduced above.
Following the architecture of Sentence-BERT~\cite{DBLP:conf/emnlp/ReimersG19},
the pooling layer takes into account the entity-level representations in addition to the attribute-wise pooling results.
Specifically, let $\mathsf{BERT}(e_a)$ and $\mathsf{BERT}(e_b)$ be the entity representations of entities $e_a$ and $e_b$ 
(the output of the \textsf{[CLS]} tokens). The pooling output is 
$$\mathsf{BERT}(e_a) \oplus \mathsf{BERT}(e_b) \oplus {\mathsf{Pooling\_homo}}(e_a, e_b) $$
when alignment information is present. Otherwise, it returns the same output with ${\mathsf{Pooling\_homo}}$
replaced with ${\mathsf{Pooling\_heter}}$.

\section{Experiment} \label{sec:exp}

In this section, we present our experimental results on real-world datasets from Indeed.com.
Specifically, we start from unlabeled entries by applying \system to generate candidate pairs and further obtain labeled instances for training and testing.
After that, we employ the two proposed model architectures to perform matching and compare with state-of-the-art EM methods.

\subsection{Experiment Setup}\label{subsec:setup}

\subsubsection{Dataset Creation}

We apply the \system pipeline to perform the job-job matching and job-resume matching tasks. 
The datasets are collected from the production pipeline of Indeed.com for the United States job market during a 6-month period from 2019 to 2020.
The obtained job postings and resumes are in JSON format (schema similar to that in Figure~\ref{fig:schema}).
The collection of job postings contains 415,656 entries after cleaning and deduplication.
For the resume data, due to privacy concerns, we used a pre-processed masked dataset created by domain experts from Indeed.com that consists of 700 high-quality resume entries to train our models. 
These two job posting and resume datasets result in a total number of 291M pairs as candidate to be matched (i.e. input of the Blocker).

We apply the blocking and labeling techniques introduced in Section~\ref{subsec:overall} to create annotated datasets for the two matching tasks. 
We follow previous work~\cite{DBLP:journals/pvldb/KondaDCDABLPZNP16,DBLP:conf/www/PrimpeliPB19} to obtain golden standard entries for evaluation that are manually labeled by domain experts.
There are 5,398 pairs for the job-job matching and 2,000 pairs for the job-resume matching tasks, respectively.
The detailed statistics of the two annotated datasets are shown in Table~\ref{tbl:datastat}.
For both tasks, we use $60$\% of the whole dataset for training, 20\% as the validation set, and 20\% as the test set. The proportion of positive and negative instances are kept balanced.

\begin{table}[!t]
	\caption{\small Statistics of the annotated datasets. 
	Note that tokenizers of pre-trained LMs may convert a word to 
	multiple sub-word tokens. A direct use of LM may
	cause a significant information loss.}\label{tbl:datastat}
	\centering
	\begin{tabular}{lcc}\hline
		\toprule
		Matching Task & Job-Postings & Job-Resume \\
		\midrule
		Cardinality & 5,398 & 2,000 \\
		\# Positive instances & 1,877 & 600  \\
		Avg. \# words per document & 406.7 & 452.9/274.6 \\
		\bottomrule
	\end{tabular}\vspace{-1em}
\end{table}

\subsubsection{Baseline Methods}

To show the effectiveness of the proposed two model architectures, we extend the state-of-the-art EM approaches for structured data to our problem that requires matching on semi-structured and document-level data collections.
Specifically, we select two methods based on classic ML and three deep learning based approaches.

\smallskip
\noindent\textbf{Logistic Regression (LR)} and \textbf{Supported Vector Machines (SVM)} are two popular classic machine learning methods. 
To apply them to our task, we first flatten and concatenate the pairs of JSON entries. 
We then train these two models using the bag-of-words and bi-gram representation 
as features.

\smallskip
\noindent\textbf{\dm}~\cite{DBLP:conf/sigmod/MudgalLRDPKDAR18} is an entity matching framework that uses Siamese RNN networks as the basic structure to aggregate the attribute values and then align the aggregated representations of the attributes. 
We report the results of its hybrid model which shows the best performance as shown in the original paper.

\smallskip
\noindent\textbf{\ditto}~\cite{ditto2021} is a recently proposed entity matching method based on pre-trained LMs. 
It combines the pre-trained LMs with data augmentation techniques for entity matching.
Note that another pre-trained LM based work proposed in~\cite{DBLP:conf/edbt/BrunnerS20} is equivalent to the baseline version of \ditto. 
Thus we exclude it from comparison.

\smallskip
\noindent\textbf{\longf}~\cite{DBLP:journals/corr/abs-2004-05150} is a recently proposed variant of Transformer for long documents that can support an input length of 16,384 much more than the 512-token limit of BERT. 
To feed the entity pairs to the model as input, we concatenate the pair of documents in each instance separated by its separator token.

\subsubsection{Environment and Settings}

All the experiments are conducted on a p3.8xlarge AWS EC2 machine with 4 V100 GPUs (1 GPU per run).
The pre-trained LMs are from the Transformers library~\cite{DBLP:conf/emnlp/WolfDSCDMCRLFDS20} and here we use BERT~\cite{DBLP:conf/naacl/DevlinCLT19} as the encoder for the two proposed  architectures and use the base uncased variant of each model in all our experiments. 
We obtain the source code of the baseline EM methods from 
the open-sourced repositories of the original papers.
For \ditto and \dm, we tune the hyper-parameters following the original setup.
To make a fair comparison, we select BERT as the encoder of \ditto.
For \longf, we set the max sequence length as 1000 with batch size 8 and learning rate $3.0 \times 10^{-6}$ due to our hardware limitation.

We use Adam~\cite{DBLP:journals/corr/KingmaB14} as the optimizer for training and fix the batch size to be 16.
We tune the hyper-parameters by doing a grid search and select the one with the best performance.
Specifically, the learning rate is selected from \{$10^{-5}$, $3.0 \times 10^{-5}$, $5.0 \times 10^{-5}$\}; 
the maximum sequence length is selected from \{128, 256, 384, 512\}; 
the number of training epochs is selected from \{20, 30, 40\}.
We use the $F_1$ score as the main evaluation metric and also report the values of precision and recall. 
For each run of experiments, we select the epoch with the highest $F_1$ on validation set and report results on test set.

%\vspace{-1em}
\begin{table}[!t]
	\centering
	\caption{Main Results (P: Precision; R: Recall; F: $F_1$ Score)}\label{tbl:res}
	\scalebox{0.9}{
		\begin{tabular}{l|ccc|ccc}
			\toprule
			 & \multicolumn{3}{c|}{Job Postings} & \multicolumn{3}{c}{Job-Resume} \\
			 & P & R & F & P & R & F \\
			\midrule
			\textsf{LR} & 0.577& 0.462 & 0.513 & 0.538 & 0.556 & 0.547  \\
			\textsf{SVM} & 0.559 & 0.504 & 0.530 & 0.571 & 0.534 & 0.552 \\
			\dm & 0.630 & 0.545 & 0.585 & - & - & - \\
			\ditto & 0.531 & 0.590 & 0.600 & 0.560 & 0.817 & 0.664 \\
			\longf & 0.521 & 0.619 & 0.566 & 0.657 & 0.614 & 0.635 \\
			\midrule
			\textsf{\system-Sequenced} & 0.582 & 0.872 & 0.698 & 0.773 & 0.908 & \textbf{0.835} \\
			\textsf{\system-Siamese}  & 0.604 & 0.855 & \textbf{0.708} & 0.626& 0.804 & 0.704 \\
			\bottomrule
		\end{tabular}
	}\vspace{-0.5em}
\end{table}

\setlength{\tabcolsep}{3.5pt}
\begin{table*}[!t]
	\centering
	\caption{Ablation analysis. }\label{tbl:ablation}
	\scalebox{0.9}{
		\begin{tabular}{l|ccc|ccc|ccc|ccc}
			\toprule
			Model & \multicolumn{6}{c|}{Job Posting} & \multicolumn{6}{c}{Job-Resume}\\
			&   & Sequenced & &  & Siamese & &  & Sequenced & & & Siamese &  \\
			& P & R & F & P & R & F & P & R & F & P & R & F \\
			\midrule
			Full Model & 0.582 & 0.872 & 0.698 & 0.604 & 0.855 & 0.708 & 0.773 & 0.908 & 0.835 & 0.626 & 0.804 & 0.704\\
			w/ rule-based classifier, w/ stru. pooling& 0.605 & 0.788 & 0.685 & 0.525 & 0.846 & 0.648 & 0.716 & 0.911 & 0.802 & 0.643 & 0.710 & 0.675 \\
			 w/ knowledge, w/o stru. pooling & 0.605 & 0.736 & 0.664 & 0.568 & 0.534 & 0.607 & 0.656 & 0.841 & 0.737 & 0.594 & 0.677 & 0.633 \\
			w/ rule-based classifier, w/o stru. pooling & 0.616 & 0.577 & 0.596 & 0.554 & 0.626 & 0.588 & 0.631 & 0.855 & 0.726 & 0.562 & 0.642 & 0.599 \\
			w/o knowledge  & 0.649 & 0.740 & 0.578 & 0.508 & 0.640 & 0.567 & 0.542 & 0.638 & 0.586 & 0.608 & 0.774 & 0.681 \\
			\bottomrule
		\end{tabular}
	}\vspace{-0.5em}
\end{table*}

\subsection{Overall Performance}\label{subsec:stoa} % should be sota

The results of comparing with state-of-the-art methods are shown in Table~\ref{tbl:res}.
Note that \dm does not support matching between two collections with different schema, thus we only compared it for the job-job matching task.
We can see that our two proposed models significantly outperform existing methods in both tasks.
In the job-job matching task, \textsf{\system-Siamese} outperforms the best method \ditto by 10.8\% in $F_1$. In the job-resume matching task, \textsf{\system-Sequenced} has a performance gain up to 17.1\%. 
The improvements are likely due to both the structural and alignment information learned by the proposed architectures.

From the above results, we further have the following observations: 
Firstly, the absolute performance numbers of the job-job matching are not as good as in the job-resume matching task across all methods.
The reason could be that the involved job titles in the postings are too sparse with insufficient instances in the training data.
As a result, it is difficult for the model to learn enough effective features 
for each job title to match correctly.
Meanwhile, the patterns of job-resume matching might be relatively simpler and 
can be captured with fewer training instances.
Secondly, we can see that the document-level model \longf does not perform well in both tasks.
Although the sparse attention mechanism of \longf can handle longer input, it remains memory-inefficient so that it limits the choices of hyper-parameters (e.g., batch size) to get better results.
Thirdly, in job-job matching task, \dm performs closely with the transformer-based method \ditto.
The reason is that compared with those datasets evaluated in the \ditto paper~\cite{ditto2021}, the text descriptions here are much longer.
Therefore, \ditto has to truncate much useful information to satisfy the restriction of max sequence length 512 of the BERT encoder.
Finally, \textsf{\system-Siamese} performs better in the job-job matching while \textsf{\system-Sequenced} performs better in the job-resume matching task.
The reason might be that even after removing irrelevant texts with techniques introduced in Section~\ref{subsec:classifier}, the sequences are still too long to keep all essential information.
In this case, the Siamese Architecture can keep 2x more information after truncation and thus produce better results. 
At the same time, the sequence length for the job-resume matching task is much shorter so \textsf{\system-Sequenced} can do a better job at capturing the interactions among tokens and thus has better performance.

\subsection{Ablation Study}\label{subsec:ablation}

We conduct an extensive ablation study to test the effectiveness of each component in the proposed method.
We propose 4 alternative solutions by 
(i) replacing the knowledge generated by the DistilBERT classifier with 
a less accurate rule-based approach (e.g., keyword filtering),
(ii) removing the structure-aware pooling layer,
(iii) performing (i) and (ii) jointly, and 
(iv) removing the injected knowledge entirely.
The results in Table~\ref{tbl:ablation} show that all proposed techniques are indeed useful to improve the performance on the matching tasks.
For example, we can see that for the job-job matching task, removing the structure-aware pooling layer will lead to 4-5\% performance loss.
Meanwhile, removing knowledge injection and applying the two model architectures on the raw documents could result in as much as 24.6\% performance loss.
The weakened domain knowledge also can cause a 1-5\% performance loss.

One observation from the results is that the structure-aware pooling layer does not contribute to an improvement for \textsf{\system-Siamese} in the job-resume matching task.
The reason could be that the Siamese network itself is not good at learning the interaction between different kinds of entries, e.g. job posting and resume.
As a result, the alignment between attributes would also not tend to provide meaningful information.
We also observe a similar case for the rule-based approach for knowledge injection:  while it provides useful structural information, it also might take away some sentences that are useful for matching the resumes.
While this can be compensated by the self-attention weights learned by \textsf{\system-Sequenced}, the performance of \textsf{\system-Siamese} becomes even worse than that without knowledge.
This result implies that we need to carefully select the model architectures for different tasks.
\section{Case Studies} \label{sec:casestudy}

In this section, we present two real case studies showing (i) how the structural/domain knowledge helps 
the developers achieve finer-grained matching by identifying and targeting relevant sections
and (ii) how \system uses the self-attention mechanism and external knowledge to explain the matching results.

\subsection{Controlled fine-grained matching}\label{subsec:idk}

\begin{figure}[!t]
    \centering
    \includegraphics[width=0.48\textwidth]{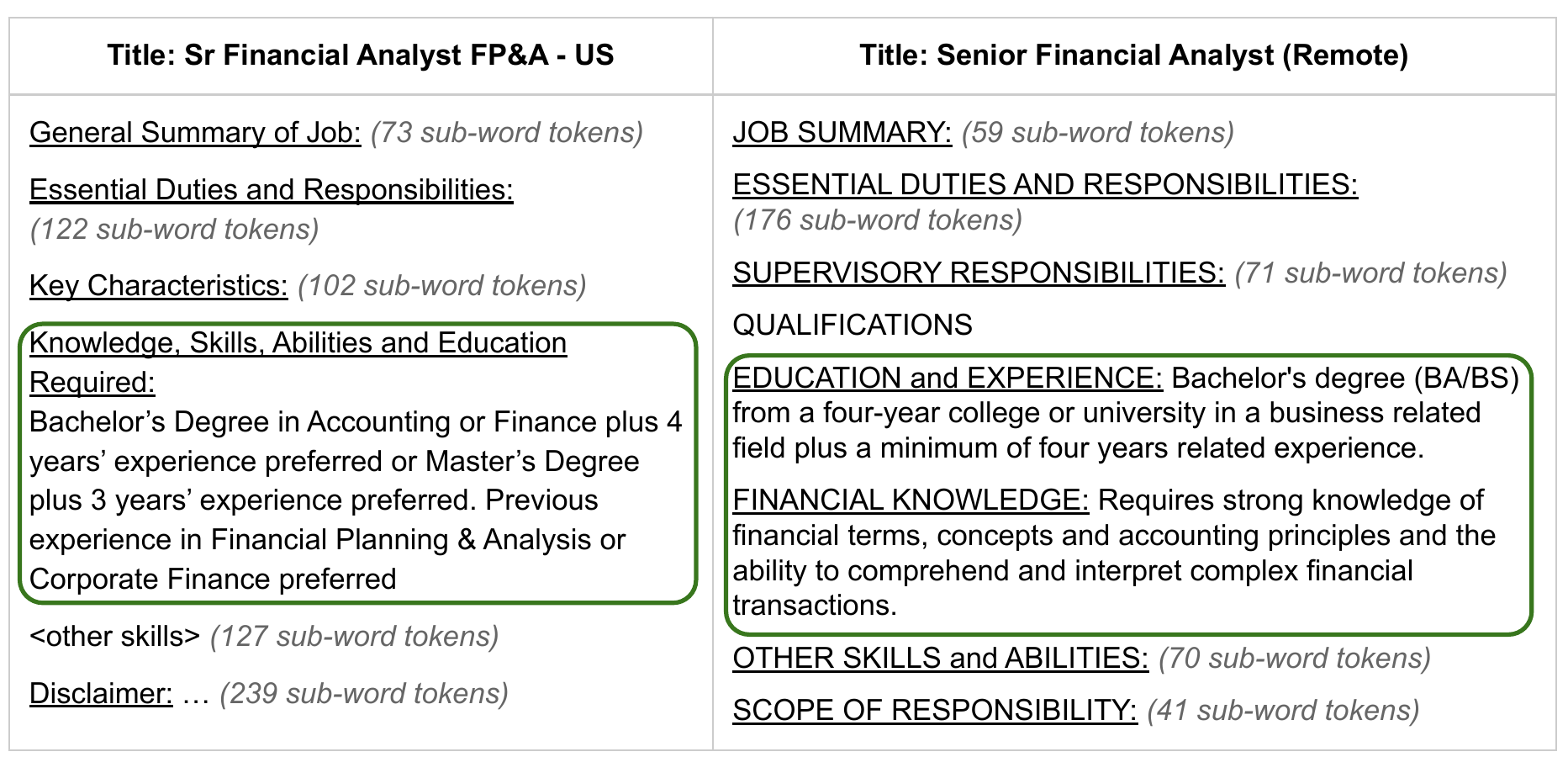}
    \caption{\small An example where the \system\ matcher predicts correctly (True) but \ditto does not.
    Long irrelevant text is replaced with word counts by the BERT tokenizer.
    While the decisive information (education and yrs. of experiences, in \green{green}) 
    gets truncated by the baseline, 
    the external knowledge classifies the text under ``qualification'' allowing \system to preserve and predict based on the text.}
    \label{fig:casestudy}\vspace{-1em}
\end{figure}

Figure \ref{fig:casestudy} shows an example from the job-job matching task where \system predicts correctly but the baseline does not.
The example consists of two matched financial analyst jobs as they require identical education and years of finance experience. 
However, the \ditto baseline failed to match them because in both entries the decisive parts of information 
are preceded by large pieces of irrelevant text (company descriptions, job duties, etc.). 
The baseline model failed to capture the key qualification information due to the 512 maximum number of sub-word tokens.
We find that this is common in the finance industry which has significantly longer job descriptions compared to the other domains.
On the other hand, in our proposed model injected knowledge facilitates topic classification to identify relevant sections of the document, in effect expanding and enriching the semantic structure of the source documents, to make more precise predictions.
Meanwhile, by knowing that the section is about ``qualification'' based on the injected knowledge (topic classification), our proposed model reserves this part in its input for a more precise prediction.

\subsection{Explainable Results}\label{subsec:explain}

For life-impacting applications such as job targeting, it is tremendously important to 
establish the confidence of developers and end-users in its decision-making process~\cite{DBLP:journals/pvldb/AsudehJ20a,DBLP:journals/inffus/ArrietaRSBTBGGM20}.
\system does so by explaining to them that the results are based on meaningful items 
in education, skills, and experiences instead of undesirable features such as race or gender.

\begin{figure}[!t]
	\centering
	\includegraphics[width=0.48\textwidth]{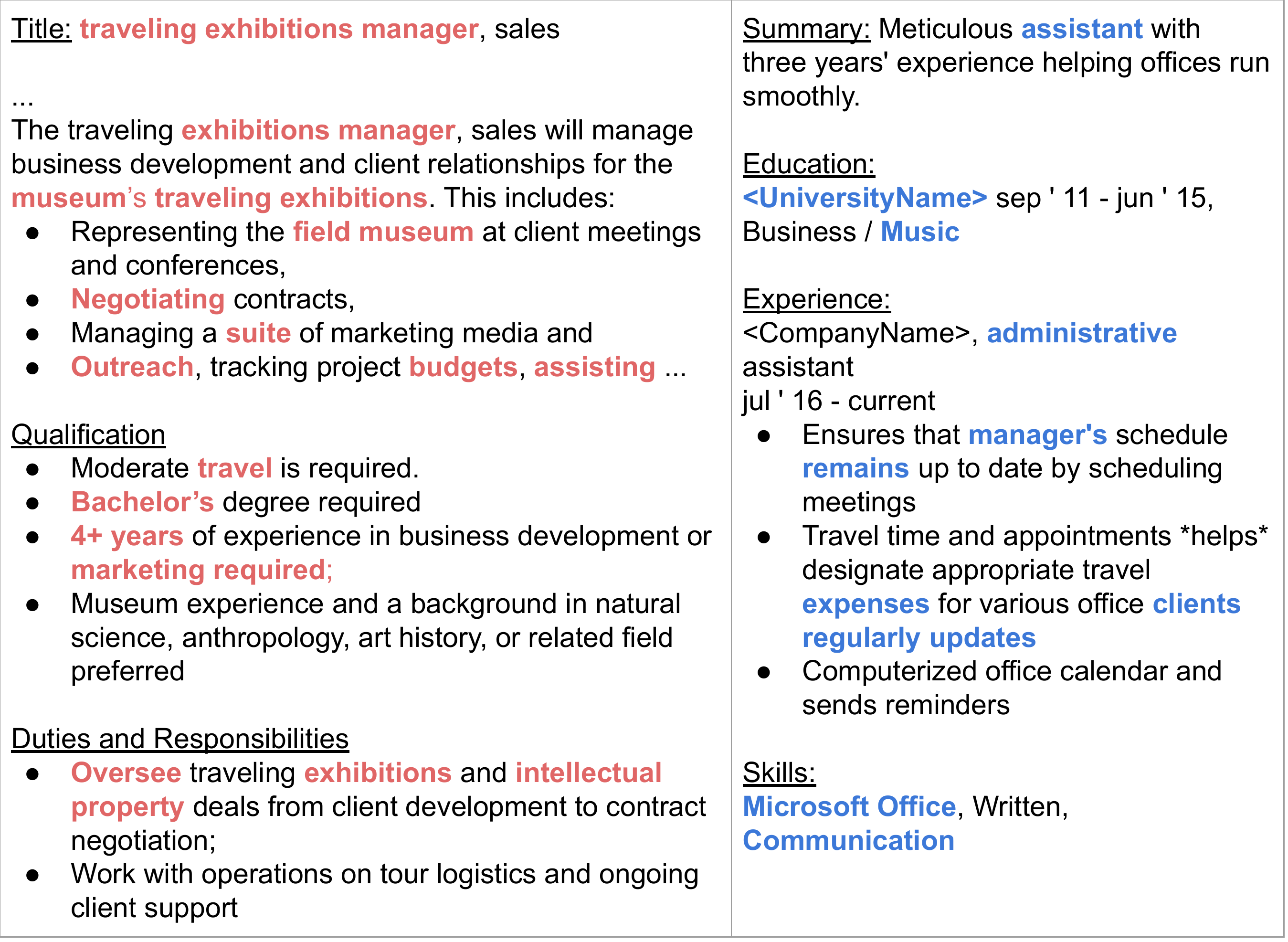}
	\caption{\small Job-resume matching with word-level explanations (highlighted in red/blue). 
		The two entries are mildly modified for privacy protection. 
		The attention weights highlighted terms related to skills and requirements while skipping less important factors such as the 
		``museum experience'' which is preferred but not required.}
	\label{fig:explanation_word}
\end{figure}

\begin{figure}[!t]
%\vspace{-1em}
	\centering
	\includegraphics[width=0.4\textwidth]{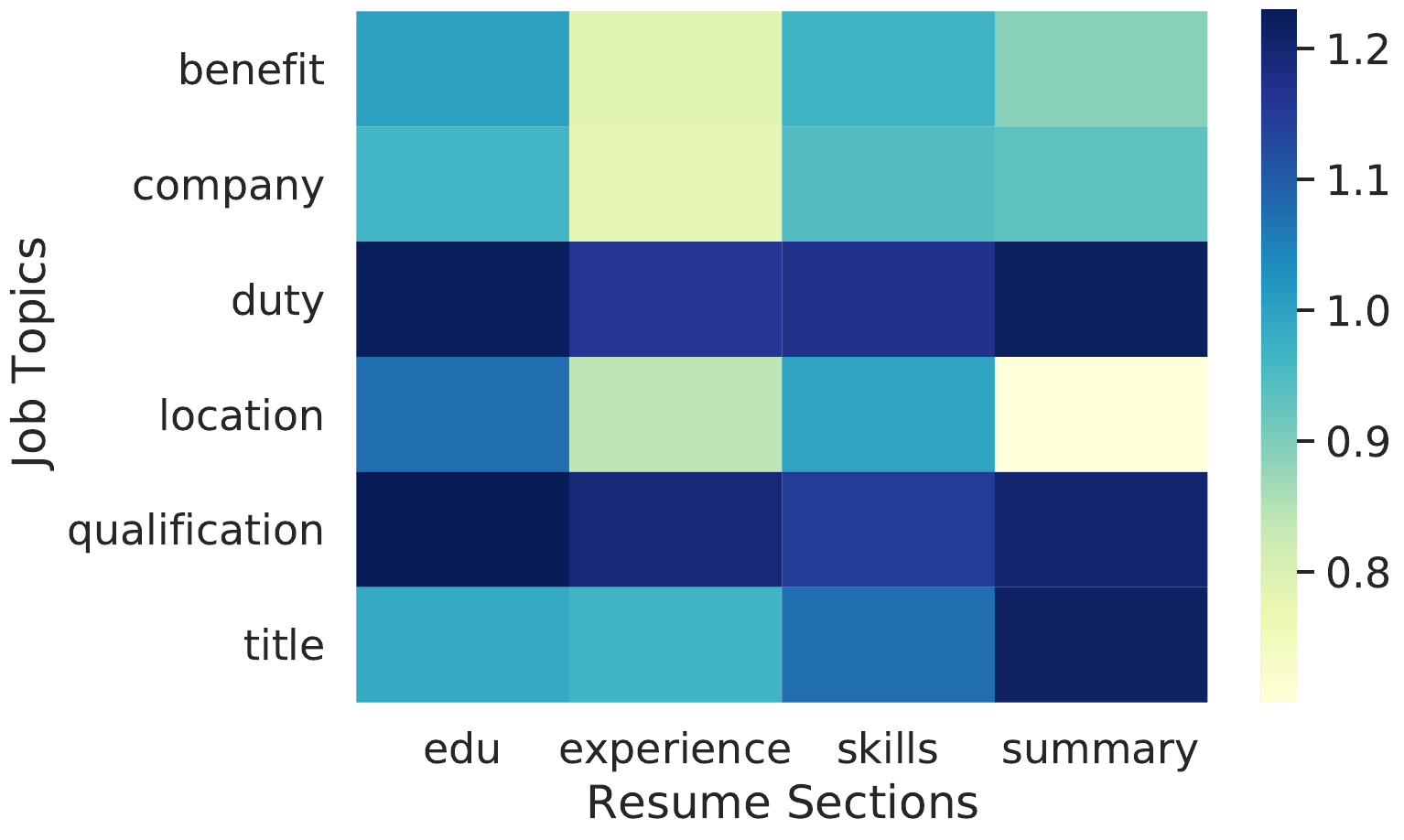}
	\caption{\small Attribute-level explanations. The ``qualification'' and ``duty'' topics of the job post 
		are deemed as more important and show the largest distance with the ``summary'' of the resume.}
	\label{fig:heatmap}%\vspace{-1em}
\end{figure}

We illustrate the explanation provided by \system in Figures~\ref{fig:explanation_word} and~\ref{fig:heatmap} with a real test example from the job-resume matching task. 
In this example, the job does not match with the candidate due to (i) the requirement for business/marketing experience and 
(ii) mismatch between the manager and the management assistant roles. 

\system can correctly predict that these entities do not match and explain the result in a fine-grained manner via \emph{attribute-level} explanations by inspecting the structure-aware pooling layer of the model. 
Recall that each attribute or structural element is encoded by BERT with a 768-dimension vector.
For job-resume matching, we denote them by $v^{\mathsf{job}}_i$ and $v^{\mathsf{res}}_j$ respectively.
We generate the explanations by computing the euclidean distance $\sqrt{(v^{\mathsf{job}}_i - v^{\mathsf{res}}_j) \cdot (v^{\mathsf{job}}_i - v^{\mathsf{res}}_j)^{\mathsf{T}}}$ for each attribute in job and resume. 
As the visualized heatmap in Figure \ref{fig:heatmap} shows, we can see that ``qualification'' and ``duty'' attributes contribute most to the mismatch result while the other attributes such as ``benefit'' are deemed as not so important.
Among the resume sections, the ``summary'' and ``skills'' attributes show a higher mismatch with the ``qualification''.

Moreover, by inspecting the self-attention weights of the LM, \system can generate \emph{word-level} explanations.
Recall that the attention output of each layer is a 3-d tensor of shape \textsf{(n\_heads, seq\_len, seq\_len)} where each element represents the weight of an attention head from a source token to a target token.
We sum over the first two dimensions and get an aggregated score for each target token.
We highlight those tokens of weights among the top-10\% in any of the first 6 layers
since these lower layers are likely to capture the lower-level word/phrase features~\cite{DBLP:conf/naacl/DevlinCLT19}. 
We visualized the results in Figure~\ref{fig:explanation_word}.
The attention weights clearly highlight the words/phrases that are position-related (e.g., ``exhibitions manager'' and ``assistant''), duty-related (e.g., ``travel'', ``museum'', ``budgets'' and ``expenses''), or qualification-related (e.g., ``Bachelor's'', ``marketing'' and ``Microsoft Office''), which are all features meaningful to the unmatch decision.

\section{Related Work} \label{sec:related}

\subsection{Entity Matching}

There is a long line of studies about Entity Matching in the database community.
Many previous studies aim at developing effective matching strategies, including syntactic similarity~\cite{DBLP:journals/tkde/Christen12,DBLP:conf/cikm/LuLW019,DBLP:conf/icde/WuZWLFX19,DBLP:conf/icde/WangLZ19} and machine learning approaches~\cite{DBLP:conf/sigmod/DasCDNKDARP17,TOT19,DBLP:conf/acl/KasaiQGLP19,DBLP:conf/sigmod/WuCSCT20,DBLP:conf/sigmod/Vretinaris0EQO21}.
Recently deep learning methods have been widely adopted in Entity Matching and achieve very promising results.
\dm~\cite{DBLP:conf/sigmod/MudgalLRDPKDAR18}, \textsf{DeepER}~\cite{DBLP:journals/pvldb/EbraheemTJOT18}, and \textsf{MPM}~\cite{DBLP:conf/ijcai/FuHSCZWK19} employ variants of the Siamese RNN network as the basic architecture.
Another two deep learning based methods \textsf{Seq2SeqMatcher} ~\cite{DBLP:conf/cikm/NieHHSCZWK19} and \textsf{HierMatcher}~\cite{DBLP:conf/ijcai/FuHHS20} improved the performance of matching between heterogeneous structured tables by applying additional alignment layers. 
Some recent studies~\cite{DBLP:conf/edbt/BrunnerS20,DBLP:conf/vldb/PeetersBG20,DBLP:conf/cikm/Wang0H21} further adopted the pre-trained LMs for the EM task.
Moreover, the data augmentation techniques have also been successfully applied to the EM problem~\cite{DBLP:conf/sigmod/Miao0021,DBLP:conf/www/MiaoLWT20}.
\ditto~\cite{ditto2021} combined the pre-trained LM with data augmentation and achieved the state-of-the-art performance on many EM benchmark tasks.

Although the presented work falls into this category of studies, there are two main differences between \system and existing works:
(i) The goal of \system is to build an end-to-end EM workflow instead of improving a single component such as blocking or matching; 
(ii) While existing studies mainly focused on matching between structured data, \system supports matching between 
data entries with different data formats such as JSON and long documents as well as based on customized definition of matching.

\subsection{Pre-trained Language Models}

Recently pre-trained LMs have become the dominant architecture in NLP applications.
They enable researchers to perform fine-tuning on training data for target tasks, which have shown groundbreaking improvements on various NLP tasks.
BERT~\cite{DBLP:conf/naacl/DevlinCLT19} adopts a Transformer-based architecture  consisting of a stack of self-attention layers to calculate distributed representations based on the similarity against all tokens.
SentenceBert~\cite{DBLP:conf/emnlp/ReimersG19} proposes the Siamese Bert architecture as well as new pooling techniques.
Other BERT variants include DistilBert~\cite{DBLP:journals/corr/abs-1910-01108}, RoBERTa~\cite{DBLP:journals/corr/abs-1907-11692}, ALBERT~\cite{DBLP:conf/iclr/LanCGGSS20} and new variants for long documents, i.e. Longformer~\cite{DBLP:journals/corr/abs-2004-05150} and Reformer~\cite{DBLP:conf/iclr/KitaevKL20}.

\subsection{Job Matching and Targeting Applications}

Job matching/targeting is an important application in recruitment platforms.
Earlier studies formalized the problem as recommendation or semantic matching problems.
LinkedIn proposed the Job Search~\cite{DBLP:conf/kdd/LiAHS16} and Job2Skills~\cite{DBLP:conf/kdd/ShiYGH20} systems to support real applications of job matching with text mining techniques.
Recent studies focused on improving the job-resume matching with different signals, such as interview history~\cite{DBLP:conf/kdd/YanLSZZ019,DBLP:conf/cikm/JiangYWXL20}, clicking rates on postings~\cite{DBLP:conf/cikm/LeHSZ0019} and browsing history~\cite{DBLP:conf/cikm/Bian0ZZHSZW20}.

While the above studies aim at proposing models to learn the contextual features, e.g. interview results of candidates, clicking and browsing on job postings to improve the matching results, our work has a more targeted problem definition, i.e. we primarily focus on texts in the job postings and resumes.
Therefore, our work is orthogonal and can be integrated into those frameworks to provide better text-based matching features.
It is a promising future work to integrate \system with rich but noisy signals such as user clicks.

\section{Conclusion} \label{sec:conclude}

In this paper, we formulate the Generalized Entity Matching problem to satisfy the need of many real applications and develop the \system framework as the solution.
\system enables the creation of a matching pipeline from scratch step-by-step.
We propose two new model architectures for the \system matcher based on pre-trained LMs.
The key novelty of the matcher lies in domain and structural knowledge injection and a structure-aware pooling layer that allows the models to
better capture essential signals for matching.
Our experiments and case studies show that \system provides significant performance gain as well as multiple granularities of explanations on the matching results.

\bibliographystyle{abbrv}
\bibliography{main,job}

\end{document}